# High-speed ultrasound imaging of bubbly flows and shear waves in soft matter

Juan Manuel Roselló,[1,2] Saber Izak Ghasemian,[2] and Claus-Dieter Ohl[2]

[1]*Faculty of Mechanical Engineering, University of Ljubljana, Aškerčeva 6, 1000 Ljubljana, Slovenia.*

[2]*Otto von Guericke University Magdeburg, Institute of Physics, Universitätsplatz 2, 39106 Magdeburg, Germany.*

(*Corresponding author: jrossello.research@gmail.com)

(Dated: December 4, 2023)

In this methods paper, we explore the capabilities of high-speed ultrasound imaging (USI) to study fast varying and complex multi-phase structures in liquids and soft materials. Specifically, we assess the advantages and the limitations of this imaging technique through three distinct experiments involving rapid dynamics: the flow induced by a liquid jet, the dissolution of sub-micron bubbles in water, and the propagation of shear waves in a soft elastic material. The phenomena were simultaneously characterized using optical microscopy and USI with bubbles as contrast agents.

In water, we use compounded high-speed USI for tracking a multi-phase flow produced by a jetting bubble diving into a liquid pool at speeds around 20 m/s. These types of jets are produced by focusing a single laser pulse below the liquid surface. Upon breakup, they create a bubbly flow that exhibits high reflectivity to the ultrasound signal, enabling the visualization of the subsequent complex turbulent flow. In a second experiment, we demonstrate the potential of USI for recording the stability and diffusive shrinkage of micro- and nanobubbles in water that could not be optically resolved.

Puncturing an elastic material with a liquid jet creates shear waves that can be utilized for elastography measurements. We analysed the shape and speed of shear waves produced by different types of jetting bubbles in industrial gelatin. The wave characteristics were simultaneously determined by implementing particle velocimetry in optical and ultrasound measurements. For the latter, we employed a novel method to create homogeneously distributed micro- and nanobubbles in gelatin by illuminating it with a collimated laser beam.

## I. INTRODUCTION

The detection and characterization of multi-phase flows in liquids or elastic solids can be intrinsically challenging when dealing with opaque media. This difficulty is further amplified when rapid dynamics are involved[1], as is often observed in industrial pipelines with cavitating flows, turbines[2], Venturi tubes[3], boiling liquids in industrial heat exchangers[4]. A similar scenario is encountered when studying fluid circulation within biological systems[5] or examining deformations in soft materials, such as those occurring during the propagation of waves through them[6]. Consequently, visualizing flow patterns and waves in these situations holds significant relevance across a wide spectrum of medical applications, including prenatal care, abdominal and breast scans, urology, and the diagnosis of musculoskeletal injuries. A significant improvement in medical ultrasound imaging (USI) was achieved with the introduction of microbubbles to enhance the contrast of USI. These ultrasound contrast agents (UCA) allowed detailed visualization of blood vessels and other body cavities, such as the urinary bladder[7]. A further refinement of this technique occurred with the introduction of stabilized bulk nanobubbles (SBNB), currently used in tumor-targeted therapy[8–10]. One of the primary advantages of SBNB over other ultrasound contrast agents is that, due to their reduced size, nanobubbles may permeate extravascular tissues and intercellular spaces. In such cases, these bubbles are typically prepared with a phospholipid coating and/or specific gases at their core to prevent dissolution once injected into the bloodstream (i.e., ensuring 'in vivo' stability). Additionally, USI plays a crucial role in industry for monitoring the operation of thermo-hydraulic machinery and gas detection in the refrigeration loops of large industrial compounds, similar to those found in nuclear facilities or chemical processing plants[11,12].

In this context, ultrasound imaging (USI) has been widely adopted as a non-invasive, affordable technique for imaging optically inaccessible mediums. One effective method to obtain information about the velocity and direction of the flow of a liquid is given by Doppler imaging[1], although the spatial resolution provided by this technique is rather limited. An opposing example would be line-by-line B-mode ultrasound imaging techniques[13], which are conventionally known for generating high-quality ultrasound images, but in contrast to Doppler imaging, their inherent limitation lies in the low achievable frame rates, making them unsuitable for capturing rapid dynamic processes. To enhance imaging speed, the plane wave imaging technique has been introduced, providing the highest achievable ultrasound imaging frame rate, limited only by the speed of sound in the medium and the imaging depth. There, a single plane wave is transmitted and the back-scattered signal from high contrast acoustic objects is recorded, allowing framing rates of several tens of thousands images[14]. For objects of weaker acoustic contrast the image quality often falls short and makes the single plane wave imaging unusable. To address the limitations of plane wave imaging, Montaldo *et al.* proposed plane wave compounding, aiming to enhance the quality of plane wave imaging while maintaining relatively high frame rates[15]. There, different plane waves with different tilted angles insonify the region of interest and the received signals after beam-forming are added together for reconstructing a higher quality USI. This way plane wave compounding can achieve signal-to-noise ratios comparable to those of traditional line-by-line imaging, but at significantly higher frame rates.

Besides acquisition of the scattered acoustic waves, ultrasound image requires a specific processing of the signals. One of the most conventional algorithms employed in the USI reconstruction is the Delay And Sum (DAS) algorithm[13,16]. In this approach, the data captured by each sensor channel undergo a two-step process: initially, they are temporally delayed to account for the time of flight from each transducer element to the depth of interest, accounting for the medium's speed of sound to align the signals received from different transducer elements; subsequently, these delayed signals are added to formulate pixel intensities. The inherent simplicity and computational efficiency of this technique, makes it suitable for real-time ultrasound monitoring, even though it might not be optimal in terms of image quality. This was partially improved by implementing Delay Multiply And Sum (DMAS) technique that produces reconstructed images of higher quality in terms of signal-to-noise ratio and contrast. There, the channel data after aligning are multiplied to each other to use the auto-correlation of the channel data[17].

Further enhancement of USI can be achieved by determining the displacement of acoustic scatters rather than their precise positions. This involves the utilization of two prevalent methods: correlation-based approaches, like normalized cross-correlation, which identifies correlations within raw radiofrequency (RF) channel data[18], and phase-shift algorithms that rely on the beam-formed In-phase and Quadrature (IQ) data for displacement estimation, such as the Kasai 1-D autocorrelator or the Loupas 2-D autocorrelator[19].

Computing the particle displacement in high-speed USI is particularly useful for capturing the rapid dynamics of liquids and soft materials[20], such as the measurements of transient shear wave propagation or the inertial collapse of a cavity[14,21]. In recent years this technique was implemented in numerous applications in medicine, including real time monitoring of the cardiac function[22,23], analysis of vascular structures[20,24] or the detection of anomalies in biological tissue using elastography[15,25]. Yet, its utilization in the field of quantitative fluid mechanics is rather limited.

In the present work, we explore the capabilities of high-speed ultrasound imaging for studying rapid flows and deformations commonly observed in bubbly liquids and soft materials. Specifically, we have utilized USI in these three distinct scenarios: 1- Bubble jetting near the free surface of a liquid pool[26]; 2- Dissolution of sub-micron laser-induced gas bubbles[27,28]; and 3- The propagation of shear waves in a soft material, namely industrial gelatin[29–32].

## II. EXPERIMENTAL METHOD

The main objective of the experiments is the detection of static or moving bubbles in a liquid, or alternatively, track the propagation of shear waves in a soft material by analyzing the displacement of embedded bubbles. To accomplish this, we combine and connect results from high-speed optical and acoustic imaging.

The experimental method employed is summarized in Figure 1. The central part of the setup was an open rectangular cuvette with interior dimensions of 76 mm in length × 12 mm in width × 13.5 mm in height. The cuvette has transparent 1 mm glass walls on all lateral sides, and was sealed at its bottom with a thin low-density (LD) polyethylene film, in contact with a linear ultrasound probe (*ATL L7–4; Philips, Amsterdam, The Netherlands*) through a thin layer of US transmission gel. The probe is connected to a research ultrasound system (*Vantage 64LE; Verasonics, Kirkland, WA, USA*) and operated at a central frequency of 5 MHz for ultrasound imaging. The LD polyethylene film is used to seal the cuvette from the bottom without blocking the transmission and reception of ultrasound signals.

The optical videos were captured at speeds ranging between 20000 and 50000 frames per second (fps) with a *Shimadzu XPV-X2* camera in combination with a *LAOWA* 60 mm *f/2.8* macro lens. The illumination of the shadowgraphs of the cuvette's interior were done with a continuous white LED lamp *SMETec* (9000 lm) or a pulsed femtosecond laser (*FemtoLux 3, Ekspla*, $\lambda$ =515 nm, pulse duration of 230 fs). The latter was particularly useful for capturing sharp images of the smallest bubbles by adjusting the frequency of the ultra-short illumination pulses to have only a single pulse within the exposure time of the high speed camera.

In these experiments, we generated two types of bubbles using two different pulsed green lasers ($\lambda$ =532 nm; FHMW = 7 ns):

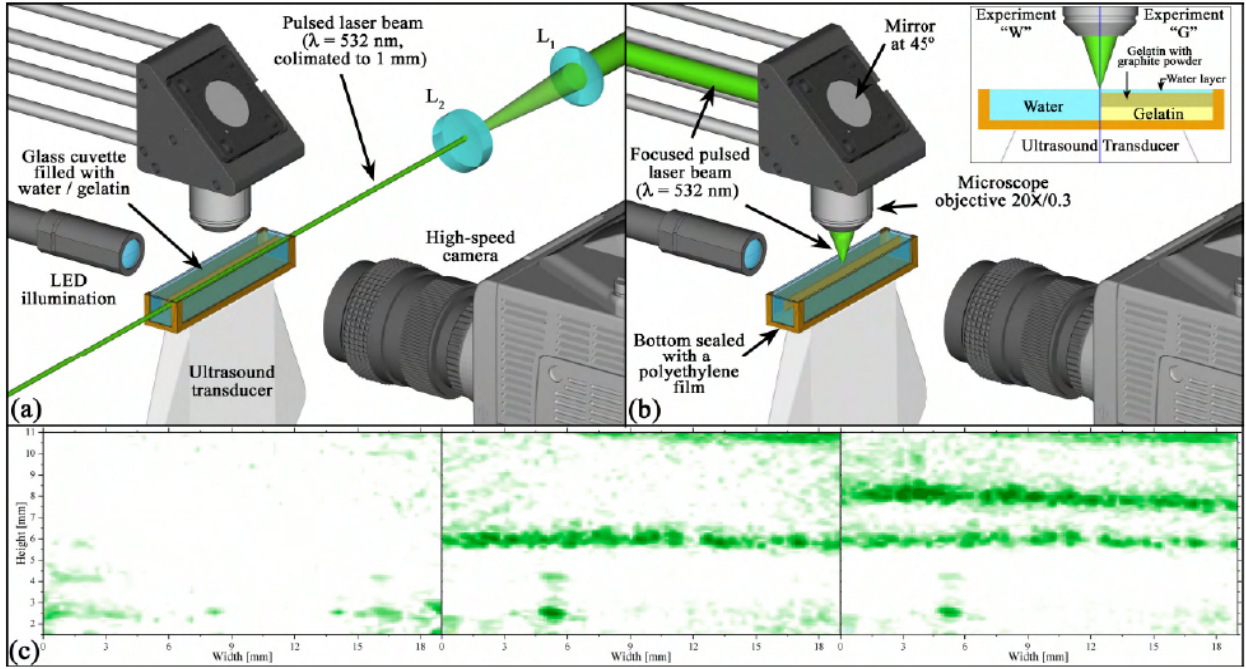

Figure 1. Experimental setup employed to investigate bulk laser bubbles (BLB), jetting cavities, and shear waves by analyzing the scattered sound signals, captured by an ultrasound transducer. These phenomena were also captured in high-speed video recordings using the backlighting technique. The central element of the setup comprised a rectangular cuvette with transparent glass walls on four of the six sides, an open-top, and a sealed bottom using a thin polyethylene film. a) The BLBs are produced when a collimated laser pulse illuminates the liquid/gelatin within the cuvette, here parallel to the surface of the ultrasound transducer. The beam diameter is shaped to 1 mm by two consecutive lenses with focal distances $L_1 = 100$ mm and $L_2 = -25$ mm, separated by approximately 75 mm. b) Alternatively, a second green laser pulse is focused just below the surface of the liquid. This leads to the formation of a cavity that evolves into a downward-directed jet. The inset illustrates the two cases studied: one with the cuvette filled with DI water (Experiment "W"), and the other conducted on a 4 % *w/w* gelatin block covered by a thin layer of water (Experiment "G"). c) US images (in false green color) depicting the inception of bubbles in a block of gelatin using a collimated laser pulse of $\sim 51$ mJ (see (a)). The three frames are taken before and after two successive laser shots at different heights inside the cuvette.

1- On one side, we produced a cluster of hundreds of *bulk laser bubbles* (BLB) with a radius below 5 µm using a *Litron Nano T-250-10* laser, following the method described in References [27,28]. Specifically, a laser beam passes through the cuvette parallel to the array of transducer elements, illuminating the volume within with a collimated laser pulse having a maximum pulse energy of approximately 51 mJ (see Figure 1(a)). Here, the Gaussian laser beam diameter is adapted by two consecutive lenses with focal distances $L_1 = 100$ mm and $L_2 = -25$ mm, spaced approximately 75 mm apart, and then clipped with a pinhole resulting in a final diameter of 1 mm.

2- Alternatively, a single larger bubble was generated by focusing a second laser pulse (*Litron*

*Nano SG-100-2*; λ =532 nm; FHMW = 6 ns, pulse energy ∼ 5 mJ) from the top using a 20× microscope objective ($f$ = 15 mm; NA = 0.3) near the free surface of the the water, as rendered in Figure 1(b). This provokes a dielectric breakdown of the liquid, resulting in the formation of a rapidly expanding cavity. As this cavity collapses, it evolves into a downward-directed jet that dives into the liquid/soft matter pool that fills the cuvette[26]. We refer to the bubbles generated using the focused laser beam as *jetting bubbles*.

In the present study, ultrasound imaging was applied to some experiments conducted in water, and others involving industrial gelatin, as detailed in the inset of Figure 1(b). In each type of experiment, different types of bubbles were used for distinct specific purposes, as explained below:

- Experiment type "W": High-speed ultrasound imaging (USI) is employed to track the bubbly flow, i.e. the jet, that occurs following the fragmentation of the cavitation bubble formed when a laser pulse is focused near the liquid surface (referred to as method "1"). Alternatively, we also characterize the dissolution process of bulk laser bubbles (BLBs) in a pool of deionized water (i.e. method "2") generated by illuminating the liquid with a collimated laser beam, as shown in Figure 1(a).

- Experiment type "G": Here, the cuvette is filled with industrial gelatin (*Gelatin 250 bloom, Yasin Gelatin Co., Ltd.*) with a concentration of 4% w/w. The sample consisted of a gelatin block infused with graphite particles within a central band, covered by a 1 mm water layer deposited on top of the soft material, as depicted in the inset of Figure 1(b). The solid particles, with an average size of ∼ 8 μm, were introduced into the gelatin during the curing process, which involved preparing the block in two stages. Initially, a plastic mold was employed to create an empty trench with a height of approximately ∼5.5 mm and a thickness of 1 mm within the gelatin block. Afterward, this void was filled with a second batch of gelatin infused with the particles at a concentration of around 0.15% v/v. The addition of the particles served two purposes: enhancing the accuracy of the optical image analysis and facilitating the nucleation of bulk laser bubbles within the volume containing the particles. The BLBs were randomly distributed over the cylindrical region exposed to the light beam as illustrated in Figure 1(a). The nucleation of BLB occurs when the laser energy is deposited on the solid objects and sublimates the surrounding gelatin as a consequence of the particle heating. While the addition of particles is not strictly necessary to induce bubbles in the bulk material, their presence increases the efficiency of the nucleation process and ensures that the bubble size is large enough to prevent bubbles from dissolving back into the gelatin, as commonly observed with the smallest bubbles when no particles were added.

As to populate the sample with a significant amount of BLBs, the horizontal laser was fired multiple times while displacing the cuvette height relative to the position of the collimated laser beam, as shown in Figure 1(c). The resulting bubbly region was defined as a band occupying a volume similar to that infused with particles. In this context, the BLBs function as markers for monitoring the propagation of shear waves using ultrasound probing. This can be attributed to the strong reflections of the ultrasound signal that these cavities generate, resulting in a significant improvement in the ultrasound image contrast when compared to the contrast created solely by the particles.

Once the sample preparation was finished, the energy of a single laser pulse was focused inside the water layer to create a cavity with a typical radius of approximately 500 µm (see Figure 1(b)), which upon collapse, jets towards the gelatin surface. The specific depth at which the laser was focused could be adjusted in order to produce the different jetting types, described in Refs.[26,33].

The high-speed optical videos obtained for both the jetting dynamics and the propagation of the resultant shear waves, were analyzed by implementing *particle image velocimetry* (PIV) using an open source code *MATLAB* named *PIVlab*[34,35].

The generation of the US high-speed images requires subsequent processing of the scattered acoustic signals captured by the ultrasound transducer. In this work, we used DMAS algorithm[17,36,37], implemented in the ultrasound toolbox USTB from Rodriguez-Molares *et al.*[38] in *MATLAB* which can reach higher signal-to-noise ratio comparing DAS algorithm. Back-scattered raw ultrasound data are acquired from transmission of plane waves with 3 different tilted angles from $-10°$ to $10°$. The frame rate of the reconstructed ultrasound images can be adjusted by employing either a single plane wave for image formation or by creating a compounded plane wave image using signals coming from multiple angles. The frame rate of the USI also depends on their depth. In this study, the effective achievable imaging speeds using one or three angles were within the range from 3 kfps to 18 kfps. Generally speaking, the quality of the reconstructed images is better with an increasing number of angles used at the expense of reducing frame rates. To enhance the quality of tracking of moving gas cavities in both the jets and the gelatin, we utilize the Loupas algorithm[19] on beamformed In-phase and Quadrature (IQ) data.

## III. EXPERIMENTAL RESULTS

The use of high-speed ultrasound imaging for studying the dynamics of multiphase flows and soft matter was assessed through three distinct prototypical experiments performed by following the methods outlined in Section II. In the first experiment, presented in Section III A, we demonstrate the ultrasound visualization of a jetting bubble near a free surface. There, the fast flow is characterized by tracking the gaseous cavities mixed with the liquid stream. In the second experiment, we discuss the capabilities of ultrasound probing to detect sub-micron laser-induced bubbles (see Section III B). In the third experiment in Section III C, the liquid is replaced by gelatin and the propagation of shear waves as a result of jetting is visualized with tiny bubbles embedded within.

### A. Flow characterization by tracking of the gas phase in bubbly liquids

In this section, we assess the feasibility of using ultrasound imaging techniques to track the positions of rapidly moving bubbles with sizes ranging from millimeters to nanometers. These bubbles are initially injected into a pool of water by a liquid jet surrounded by gas and generated through method "1". This multi-phase jet travels at an initial speed exceeding 20 m/s, and then fragments along the cuvette floor. To resolve the dynamics of these complex gas structures in the USI, we employed compounding plane wave imaging with three different tilted angles, namely 10°, 0°, and −10°. The beamformed IQ data was reconstructed using the DMAS algorithm[17]. Figure 2 presents optical images side-by-side with the corresponding ultrasound images of two exemplary cases of a jetting bubbles. This jetting regime has been named "bullet" jet[26], and is observed for the non-dimensional laser focus distance in the liquid of $\gamma = \frac{d}{R_{max}}$ below 0.25, where $d$ represents the bubble laser seeding depth and $R_{max}$ is the maximum radius attained by the cavity. For a detailed description of the formation mechanism behind the bullet jets, refer to References[26,33].

The frames of Figure 2a and b are bordered at the top with a free surface. The US transducer is imaging the scence from below. Figure 2(a) reveals that the US images allow for a precise tracking of the jet tip (see arrow "A"), even when it moves at an initial speed of approximately 20 m/s. The cavities are more clearly resolved in the US images when the path that separates them from the ultrasound probe is free of other reflective objects, as previously reported in the case of a single bubble[14]. However, this level of precision is not achieved in every region of the probed volume. For example, we notice a "shielding" effect coming from the distortions introduced by the cavities closer to the US transducer, which results in noisy regions farther away. Additionally, the US images exhibit a higher intensity near the free surface (as indicated by arrow "B" in Figure 2a)) where the gas cavities are rather small. A similar artifact is observed where bubbles are grouped as in the region pointed by arrow "C". This effect is found in a second run shown in Figure 2b. There, the arrows highlight how micrometer-sized bubbles, which are invisible in the optical images (with a resolution of $\sim 20$ µm/px) still reflect and produce a strong US signal. The eventual overestimation of bubble sizes is a characteristic artifact often encountered in ultrasonic imaging. This effect is observed when tiny bubbles cluster around a specific spot, creating a relatively homogeneous "cloud-like" region in the USI that occupies a larger volume than that encompassing the actual bubbles, as is clear in Figure 2b, see arrow "F".

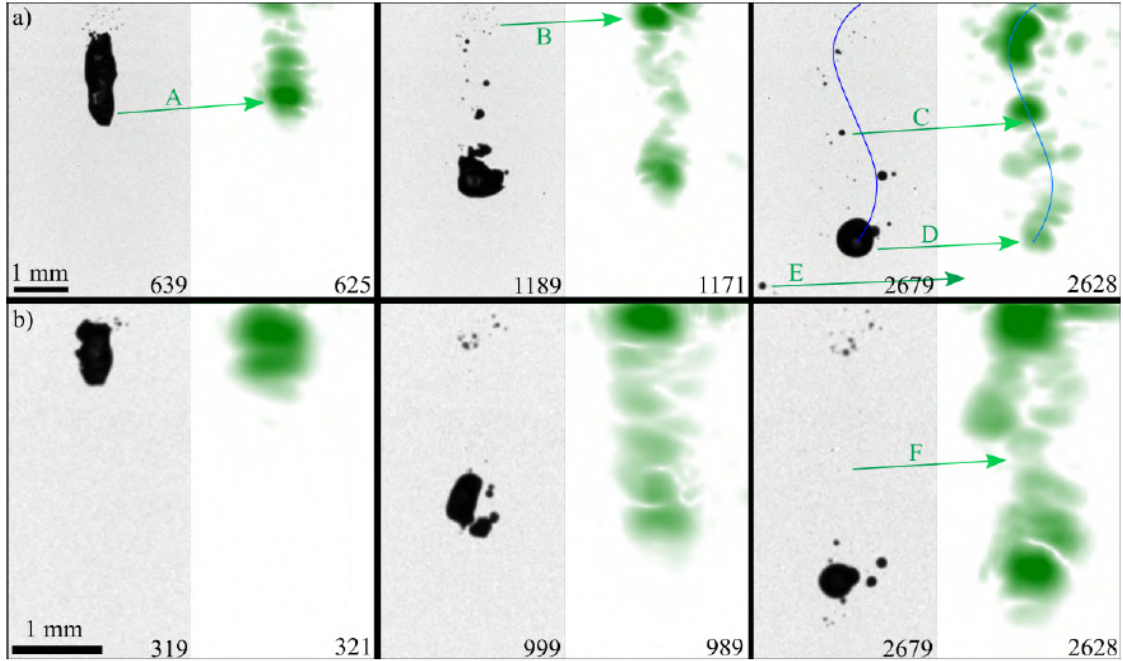

Figure 2. Optical image sequences of a bullet jet ($\gamma = 0.21 \pm 0.02$) diving into the water pool, shown alongside the corresponding ultrasound images. The optical and ultrasound images have similar times and share the same pixel size. Bubbles are generated by the focused laser pulses near the free liquid surface (at the top of the frames) and jet towards the bottom of the cuvette, where the ultrasound transducer is located (1.5 mm below the frame end). a) The jetting bubble enters the liquid at an initial speed of $\sim 10$ m/s. It progressively slows down and stops before reaching the cuvette bottom, leaving a winding trail of micron-sized gaseous cavities (indicated by the line in blue) on its path. Here, the jet tip is marked by with arrow "A". The arrows labeled "B" and "C" indicate the presence of small cavities left by the fragmentation of the gas phase surrounding the jet. Remarkably, these bubbles produce a high intensity in US image. Observing the intensity of the jet tip as it advances, it becomes evident that as the bubbles approach the transducer, they appear progressively dimmer in the reconstructed image. This trend is highlighted by the labels "D" and "E". b) The image sequence demonstrates the detection of gaseous cavities with sizes below the optical pixel resolution through ultrasound probing. An example of this phenomenon is highlighted by arrow "F".

Another limitation in USI involves the challenge of visualizing objects positioned in close proximity to the ultrasound transducer, entering the so called transducer's "dead zone"[39]. Here we observe that objects which are very close to the surface of the transducer appear dimmer than the rest. This phenomenon is noticeable in the last frame of Figure 2(a); see arrows "D", where the signal from the jet tip diminishes as it progresses towards the transducer. Furthermore, the small gas cavity indicated by arrow "E" is nearly imperceptible in the ultrasound image, indicating that it is entirely within the dead zone.

At this point, it is evident that the method fails to establish a definitive relationship between bubble size and the brightness observed in ultrasound images, potentially due to some bubbles being resonant emitters. In addition to the ultrasound shielding produced by the bubble clustering and the limitations in the spatial sensitivity of the sensor, one must consider the effect of spurious reflections on the cuvette walls and the free surface of the liquid. Despite its relatively low accuracy, the technique has proven to be highly effective in detecting the presence of a gas phase and providing insights into the dynamics of the bubble cloud. For example, it allows for an accurate measurement of the liquid/gas flow velocity and an estimation of bubble distribution over larger areas, as illustrated by the blue curve in Figure 2(a), representing the winding trail of small bubbles left after the passage of the jet.

Some of the mentioned limitations can be overcome by computing changes in the gas phase distribution instead of locating the positions of the cavities. Accordingly, we implemented a phase-shift algorithm that utilizes the beam-formed In-phase and Quadrature (IQ) data for displacement estimation, specifically, the Loupas 2-D autocorrelator[19]. Once more, the beam-formed IQ data is reconstructed through compounded imaging from three plane waves. One example of this alternative signal processing is presented in Figure 3.

After comparing the optical images with both the USI reconstructed videos (e.g., Figure 2) and the videos obtained from the displacement maps, it becomes evident that the speckle noise, commonly observed in USI[40], is considerably reduced in the displacement maps. The robustness of this alternative visualization method makes it better suited for tracking individual gas pockets and characterizing the dynamic behavior of the cluster as a whole. This is exemplified by the small group of bubbles detaching from the main stream (indicated with a black arrow) and also by the way the bubbles at the bottom of the frame change their direction after reaching the surface of the transducer (illustrated by the red arrow).

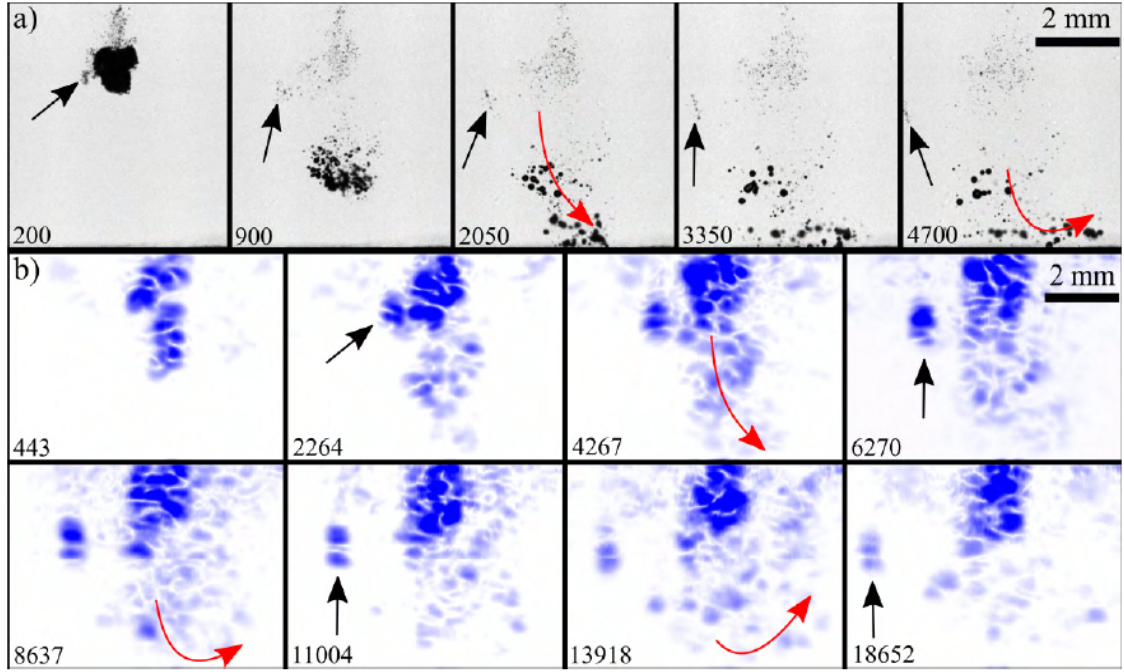

Figure 3. Displacement dynamics of the gas phase of a jetting bubble. a) Optical high-speed image sequence showing the injection of gas produced by a jetting bubble ($\gamma = 0.6 \pm 0.03$). b) Here, the movement of the bubbles is tracked with ultrasound to determine the long-term behavior of the flow induced by the jet shown in a). The gas cavities resulting from the fragmentation of the jet disperse after reaching the transducer. Eventually, they float back to the surface of the water due to buoyancy. The direction of the bubble flow is indicated by a red arrow. The black arrows follow the trajectory of a small group of bubbles that diverge from the main flow. This example demonstrates that USI allows for visualizing individual bubbles in the displacement map.

## B. Dissolution of bulk laser Micro- and Nanobubbles monitored through ultrasound imaging

One of the most intriguing results observed in the experiments presented in the previous section was the possibility to detect cavities with sizes below the resolution of the optical system, i.e. invisible in the high-speed images (e.g. Figure 2(b)).

In this section, we further explore this idea by investigating the dissolution of micro- and nanobubbles through USI. Consequently, we performed a second experiment using water as the working liquid (Experiment "W") and employed bubble generation method "2", as detailed in Figure 1(a), to create a cluster of bulk laser bubbles (BLB) homogeneously distributed over the volume illuminated by the laser beam in the water, similarly to bubble configuration shown in Figure 1(c). Specifically, we monitored the decay of ultrasound signals scattered by the bubble cluster as the gas/vapor in their interior condenses and returns to the liquid.

The primary objective of this study is to correlate the lifetime of the bubbles within the cluster with their size, as proposed in Refs.[27,28]. These previous works demonstrate that the number and the eventual size of the bubbles induced in the liquid by the collimated laser beam is proportional to the energy of the laser pulse. Figure 4 presents measurements of the temporal decay of the averaged intensity of ultrasound scattered by bubble clusters produced with three different pulse energies $E_L = 7\,\text{mJ}$, $E_L = 11\,\text{mJ}$ and $E_L = 15\,\text{mJ}$. In this figure, the curves labeled as A, B, and C represent the average of 4 to 8 measurements of the mean pixel intensity within the ROI presented in the single-plane wave USI at the top of the figure.

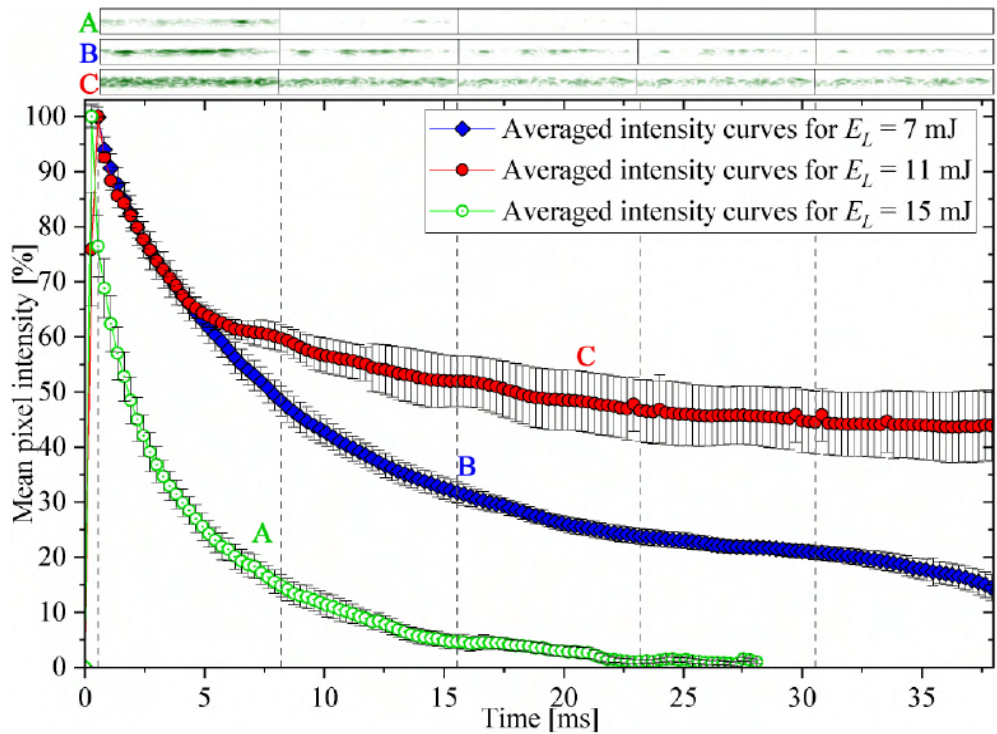

Figure 4. The dissolution dynamics of the BLBs were assessed by observing the temporal evolution of the mean pixel intensity generated by the bubbles in the USI taken at intervals of 164 µs. The decay in the intensity curves A, B and C, derived from three distinct laser pulse energy levels, is correlated with the bubble population detected by the ultrasound sensor with single-plane wave, shown in the frame sequence on top of the plot. The mean value of the noise around the ROI in the USI was subtracted from the curves before normalization. There, the vertical bars represent the standard error of the mean value. The frame width corresponds to 19.2 mm and the left edge of each frame is aligned with a specific time indicated on the horizontal axis.

The image sequences in Figure 4 provide clear evidence of the correlation between pulse energy, the density of the bubble population, and their persistence in the liquid (i.e., the bubble lifetime). This relationship is further confirmed by the corresponding normalized intensity curves. According to the Epstein-Plesset theory (E-P), the dissolution time of a gas cavity scale with its initial radius[41]. Consequently, the slower decay observed for cases with higher energy pulses necessarily implies the presence of more and larger bubbles.

The USI displayed in Figure 4 also reveals that the bubble size distribution along the "line" generated by the laser beam is not uniform, contrary to the initial assumption. For instance, in example "B", bubbles in certain sections of the line remains for a longer duration, indicating their larger size. This is because BLBs are produced when the laser heats certain impurities, such as solid particles, that are naturally suspended in the liquid. As demonstrated in Ref.[28], larger particles contribute to the formation of bigger bubbles, which are briefly visible immediately after their generation. These relatively larger bubbles, typically constituting less than 8% of the total population, result in a flattening effect on the average mean pixel intensity due to their slower dissolution rates.

Considering the latter observation, we conducted a comparison of the ultrasound intensity derived from individual measurements taken within a reduced region of interest (ROI) that matched the frame size of the high-speed optical videos. The results presented in Figure 5, further cement the connection between the intensity decay and the bubble size. The image sequences at the bottom of the figure displays the optical images for three key moments: just before the laser shot ($t = 0$), immediately after the laser shot ($t = 1\,\mu s$) and some microseconds after the laser seeding when the bubbles reach their "rest" radius ($t = 8\,\mu s$). There, the bubbles highlighted with purple circles indicate nucleated bubbles that are visible with a resolution of $6.5\,\mu m/px$ (i.e. with a radius below $\sim 3\,\mu m$).

In previous works on this type of bulk laser bubbles[27,28], the Epstein-Plesset equation was used to calculate a histogram of the bubble size distribution. This calculation was based on the temporal decay in the number of BLBs optically detected after the laser shot. In the current case, the previously discussed accuracy limitations prevent us from establishing a direct correlation between the evolution of the US intensity and the characteristics of the bubbles. One of the primary issues with the ultrasonic bubble detection method is the inability to resolve the number and size of the bubbles. This challenge arises because both a millimetric cavity and a cloud of nanobubbles distributed over a similar volume generate comparable scattered signals. In spite of that, we can perform a rough estimation of the maximum size of the BLB by measuring the dissolution time of the entire cluster for the cases "A", "B" and "C", corresponding to different laser pulse energies. In case "A" ($E_L = 7\,mJ$) 80% of the bubbles are not longer detected after $t = 3\,ms$, consistent with the size estimation performed using a different probing method in Ref.[28] but under the same laser beam configuration. In this previous study, approximately 150 bubbles were observed within the same ROI, with typical radius below 100 nm. In case "B" ($E_L = 11$ mJ) the same estimation would suggest that 80% of the bubbles have a radius below 750 nm. For the scenario with the highest pulse energy "C" ($E_L = 15$ mJ), the majority of the bubbles have a radius below 1 µm. However, there are multiple cavities which exceed that size, as indicated in the flattening of the red decay curve in Figure 5 and confirmed by the bubbles that are clearly visible in the optical frame.

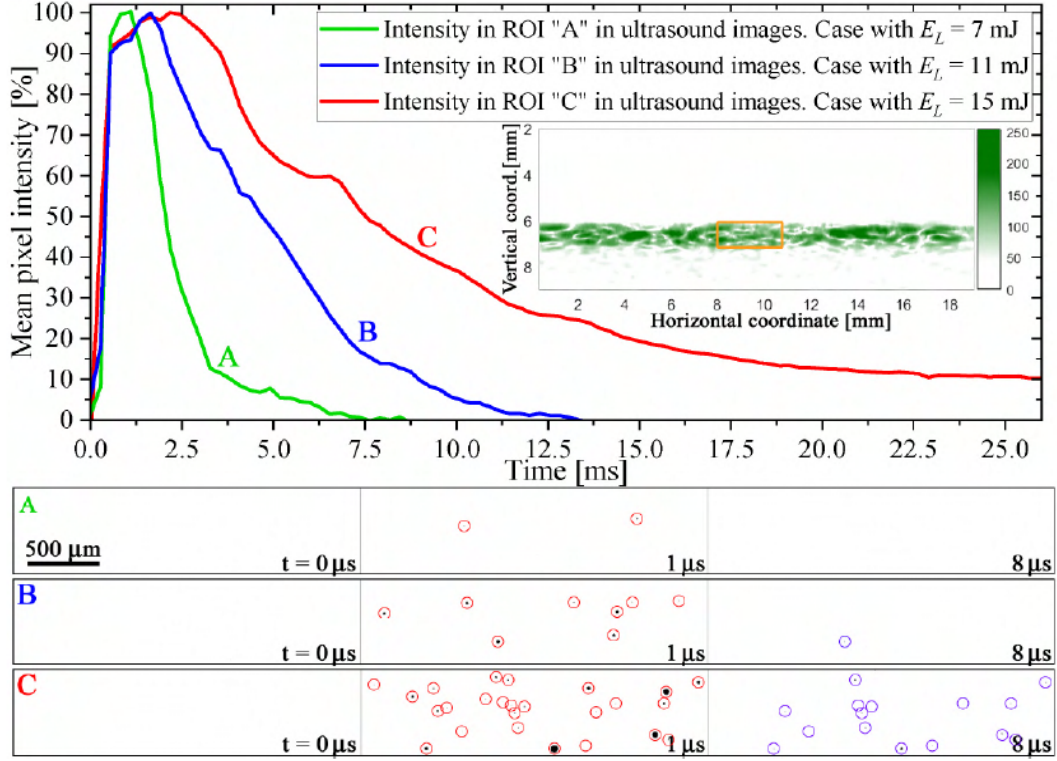

Figure 5. Comparison between the mean pixel intensity in the with single-plane wave ultrasound images (upper panel) and the characteristics of bubbles detected in the optical images (lower panel) for three different energies of the laser pulse ($E_L$). The signal intensity captured in the ultrasound images correlates with the dissolution dynamics of BLBs within the region of interest (ROI), derived from the observation window in the optical frames, as highlighted by the orange frame in the inset. Each one of the lower panels A, B and C displays three optical frames of the ROI captured at different instants: the moment of laser shot (i.e. t = 0 µs), when the laser-induced bubbles reach their maximum radius (t = 1 µs), and after the bubbles have settled at their ambient radius ($t = 8$ µs). The red and purple circles indicate the presence of a visible bubble at the different instants, showing the existing correlation between $E_L$, the number and size of the resulting bubbles and their lifetime, here indicated by the decay rate of the pixel intensity in the ultrasound images.

Estimating attributes of the gas cavity population using the E-P equation also depends on the minimum detectable bubble size with the USI. This becomes particularly relevant when dealing with bubbles that undergo rapid dissolution, as is the case with nanobubbles. In such instances, the impact of underestimating the bubble lifetime becomes increasingly significant for smaller bubbles. Therefore, it is essential to conduct a detailed characterization of the ultrasound system's capabilities for each specific experimental scenario using an alternative method, such as that described in Ref.[28], to effectively evaluate the range of applicability for the proposed US method.

On the positive side, using USI for monitoring the gas diffusive stability of bubbles can be performed in real-time with a single measurement of the physical system. This stands in contrast to the method described in Ref.[28], which requires several hundred measurements taken under highly controlled conditions and is therefore not suitable for non-repetitive phenomena.

### C. Ultrasound visualization of shear waves in gelatin

The laser seeding of homogeneously distributed bubbles has proven to generate good contrast in the USI. Particularly, the potential to create a localized bubble cluster in semitransparent media with a single laser shot represents an interesting tool. In a third experimental application utilizing USI, we make use of the bubble seeding technique to investigate the propagation of a shear wave in a soft material. For this part of the work we performed the experiment type "G" described in Figure 1. In this experiment, a block of gelatin is first covered with a 1 mm thick layer of water which is later hit by a focused laser at a given depth ($d$). The laser bubble evolves into a liquid jet that pierces the block surface and penetrates the gelatin downwards, generating shear and compressive stresses that propagate radially from the injected liquid column.

In this study, we will disregard the shock and rarefaction waves emitted during the explosive dielectric rupture produced by the focused laser on the water layer. These waves travel at a speed far beyond the temporal resolution achievable by the USI. Instead, our focus will be on the shear waves produced by the vertical displacement of the gelatin, which constitutes the predominant component of the jet-induced deformation. For the USI processing, we applied the same computational methods as in Section III A, i.e. utilizing the DMAS technique for compound image reconstruction using three angles and the Loupas 2-D auto-correlator for displacement calculation.

The gelatin was embedded with graphite particles and bubbles disposed over a vertical layer that traverses the cuvette along its length. Although BLBs dissolve quickly in water, they undergo an expansion process when produced in gelatin as a consequence of gas diffusion[42,43]. The BLBs initially had sizes that were invisible in the optical images, even at the highest resolution of around 6 μm/px (see Section III B), but they could reach a diameter of around 1 mm after a few minutes. Here, the particles serves as trackers of the material deformation in the optical PIV analysis, while the bubbles have a similar purpose as contrast agents in the USI.

In a previous study[33], we established that varying the parameter $d$ in experimental setup "G" can yield different types of jets, each resulting in distinct penetration depths of the liquid into the gelatin. This variation is a direct consequence of the particular jetting dynamics exhibited by the bubbles in each case. In the current study, we continue to employ the same jet classification and further analyze the phenomenon, demonstrating not only that the jets can generate strong shear waves in the soft material but also that their penetration depth has a significant impact on the shape and amplitude of the wave front.

In Figure 6, we compare the shear waves generated in the gelatin by a single laser induced bubble with three types of piercing jets. Therefore, the shear waves are analyzed shortly after the liquid reaches its maximum penetration using optical PIV (second column) and US (right column) particle displacement in Figure 6. The results indicate that with a stronger jet that penetrates deeper into the soft material, the amplitude of the shear waves increases, and their wave front transforms from a spherical wave to a cylindrical wave. It is emitted from the interface between the penetrating liquid and the gelatin. Interestingly, we find no effect on the speed of the jet penetration speed, which fluctuates between runs between 14 m/s and 17 m/s for the jets shown in Figure 6. Still, they significantly exceeding the shear wave propagation speed[44], typically below 1 m/s. The results in Figure 6 reveal an excellent agreement between the optical and ultrasound methods. Moreover, they show that both the maximum velocity of the particles displaced in the gelatin and the shear wave oscillation amplitude are notably higher for all the piercing jet cases, compared to the spherical shear wave propagating from a single point source produced when the bubble is seeded directly within the gelatin (e.g., Figure 6(a)).

Given that the particles and BLBs are positioned within a 1 mm thick observation plane and due to the reduced focal depth in the optical images, the material displacement measured in both the PIV analysis and the USI is minimally affected by components of the spherical/cylindrical shear wave propagating in directions outside the plane of view, revealing an excellent agreement between both methods. The PIV images also suggests that the bubbles are much smaller than the apparent "wavelength" of the deformation waves. Thus, they undergo a displacement without presenting a significant deformation in the shear wave front.

When examining the radial propagation of the shear waves, it becomes evident that their general oscillatory behavior is quite similar, irrespective of the jet type and penetration depth. This observation is supported by the temporal evolution of the material displacement obtained from the USI, depicted in Figure 7. Each curve in the plot represents the average particle speed within the region outlined in red, as shown in the examples included in Figure 6. This characterization confirms the relation between the oscillation amplitude produced by the shear waves and the penetration depth of the jets.

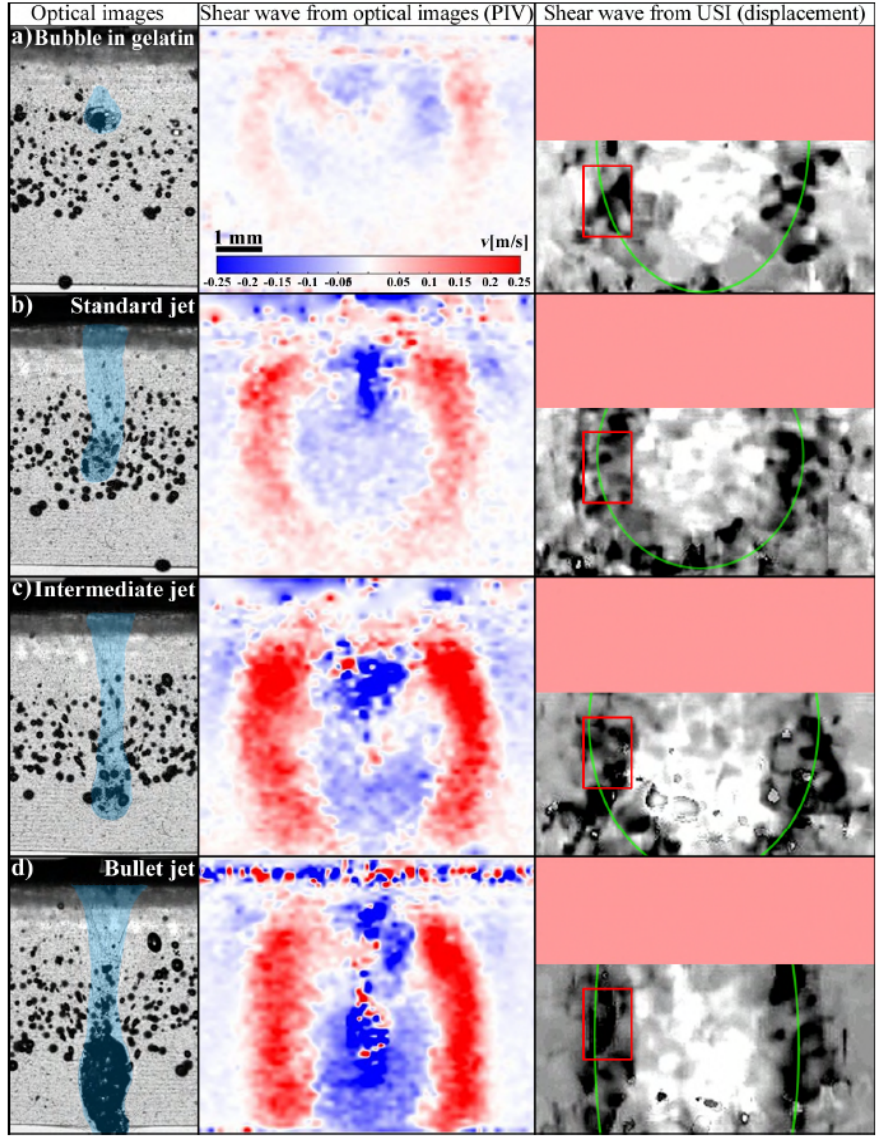

Figure 6. Simultaneous optical and ultrasound imaging of shear waves emitted by a laser bubble (a) and three distinct types of jets (b-d). Specifically, the following scenarios are presented: a) a single laser bubble produced approximately $\sim 1$ mm inside the gelatin, b) a "Standard" jet, c) an "Intermediate" jet, and d) a "Bullet" jet. Here, the first column shows an optical frame taken at the moment when the bubble/jet reaches its maximum depth. The injected liquid was highlighted with a light blue coloring. The second column presents a PIV image of the fully formed shear wave, and the third column shows an image obtained from the ultrasound displacement data. Both the PIV and the USI show the averaged vertical displacement of 3 frames centered at a time $t = 2.5 \pm 0.2$ ms. The optical images used in the PIV were captured at a frame rate of 25 kfps, whereas the USI had an effective frame rate of approximately 5.5 kfps. The contrast in the USI was arbitrarily adjusted to improve the shear wave visualization. Additionally, the surface region was excluded from the analysis to eliminate strong reflections that would mask the signal from the shear waves. The red frame marks the region analysed in Figure 7.

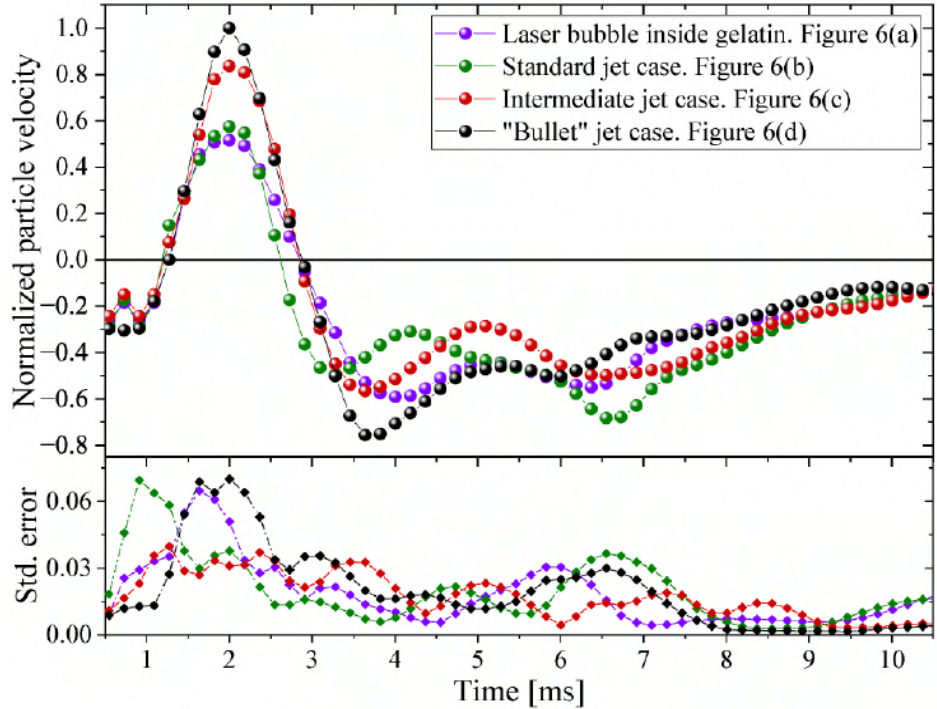

Figure 7. Temporal evolution of the US particle displacement velocity from shear waves generated by the different types of jetting bubbles shown in Figure 6. Each curve represents the average of velocity profiles within the region framed in red on the third column of that figure. The data has been normalized with the maximum displacement rate obtained for the bullet jet case and considering a positive velocity when the particles move downwards. As the different types of jets involve different timings, the temporal reference of the curves was adjusted to align the displacement peak and facilitate the comparison of the shear waves.

The similarity in the characteristics of shear waves and their speed across various jet types represents a significant advantage of this method for studying the rheological properties of an opaque material or in cases where optical access is limited. For instance, the results of elastography analysis would not present a significant variation against fluctuations in the depth where the laser is focused or the precise amount of water deposited on top of the soft material.

Figure 7 also supports the notion that bullet jets are the most effective type for generating strong and uniform wave fronts. Both factors depend on the penetration depth of the jet, which can be scaled by maintaining the ratio between the laser cavity size (controlled by the pulse energy) and its seeding depth. Considering the latter, we will focus now on the jet piercing mechanism and the details of shear wave generation for the bullet jet case only.

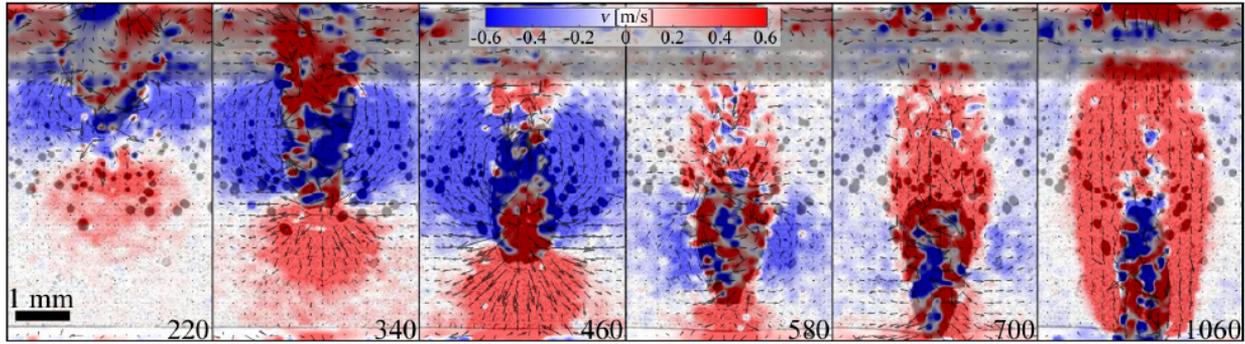

Figure 8. Details of the bullet jet piercing dynamics were evaluated through PIV. The color bar represents the vertical component of the velocity. The numbers indicate time (measured in µs) after the focused laser shot on the top liquid layer.

Both properties are investigated in greater detail in Figure 8 using PIV images computed from high-speed video captured for a bullet jet. The penetration dynamics is depicted in the first few frames of the sequence. Initially, the upper part of the gelatin is reached by the downward jet produced by the collapse of the laser cavity in the liquid phase. The impact of the liquid column on the soft material creates a high pressure point, which compresses the gelatin around the jet tip. This local overpressure displaces the gelatin, initiating the downward movement of the liquid column. More importantly, it stretches the soft material in a direction perpendicular to the jet's path (as indicated by the arrows in the lower side of the third frame in Figure 8), ultimately causing gelatin fracture and further facilitating liquid penetration. Interestingly, following the jet tip's passage through a specific section of the gelatin, the initially downward-compressed material is now displaced upwards to the upper side of the jet, as depicted in the blue-colored region of the PIV sequence. The *compression, fracture, and release* dynamics found here for the bullet jet is similarly observed for the standard and the intermediate jet shown in Fig. 6(b) and (c).

Finally, as the jet decelerates and reaches its maximum penetration depth, the confined shear stresses are emitted as shear wave[33,45] already visible in the last frames of Figure 8 (depicted in red).

Figure 9 shows an extended tracking of the wave front produced by a bullet jet, using both PIV techniques from the optical images and ultrasound high-speed imaging. Once the shear wave is emitted, the cylindrical wave front propagates into the tissue mimicking phantom (i.e. gelatin) without suffering significant distortion apart from the wave's overlap with its own reflection from the free surface, which is visible in the upper part of the final frame of both sequences depicted in Figure 9. Remarkably, the overlapping of the optical images with the PIV analysis reveals that the presence of the BLB does not distort the wave front. This is evident when examining the propagation dynamics in the middle region of the frames, which is densely populated by bubbles, and the region near the bottom where there are almost no bubbles.

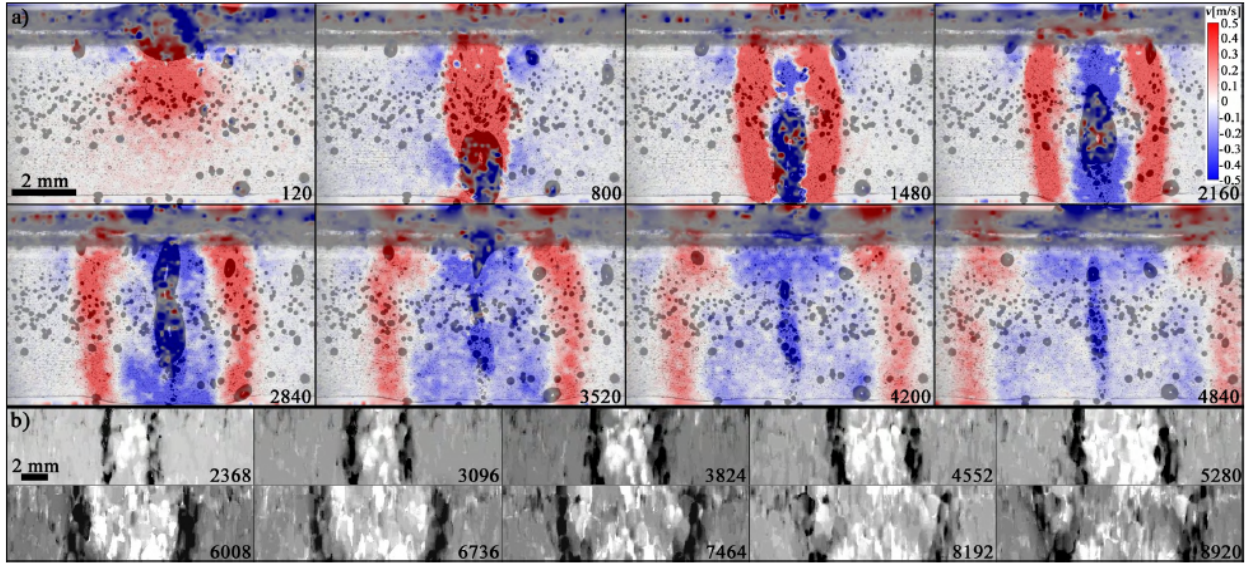

Figure 9. Generation and propagation of shear waves from a piercing "bullet" jet into gelatine. The high-speed jet creates a vertical shear stress in the soft solid which propagates as a shear wave to the sides. The gelatin was infused with graphite particles for optical tracking of the shear waves using PIV. Similarly, gas bubbles were used for wave tracking with ultrasonic imaging. a) Vertical velocity map extracted from PIV data overlaid on high-speed optical images acquired at a frame rate of 25 kfps. b) Corresponding ultrasound imaging of the shear waves with a frame rate of 5.5 kfps produced by applying a 2-D autocorrelator to the ultrasound data. As in Figure 6, the region near the surface was cropped from the USI.

The shear wave remains detectable across distances that cover the entire length of the US transducer (approximately 2 cm). However, considering that the bullet jets scale with the laser cavity size[26] and that larger jets would also generate stronger shear wave fronts, it is possible to further expand or reduce the volume of soft material being probed.

This method offers a way to study shear wave propagation in small volumes, allowing for "real-time" elastography analysis in highly localized regions. This capability can be useful for detecting tissue anomalies and characterizing the mechanical properties of soft materials. The elasticity modulus ($E$) from linear elastic materials can be computed from the shear wave propagation speed ($V_s$), the density ($\rho$), and Poisson's ratio ($v$) using the relation $E = 2\rho V_s^2 (1+v)$[25,46].

One method to obtain precise measurements of the propagation speed of the shear waves $V_s$ is by generating a particle velocity map, as detailed in Figure 10. These maps were created, for both the PIV data and the USI, by averaging the particle velocities ($v$) along each vertical line of pixels within a horizontal band ranging from 1.7 mm and 4.7 mm beneath the gelatine surface. The selection of the Region of Interest (ROI) at this depth range was based on the horizontal propagation direction of the wave fronts for all jet types (refer to Figure 6), making it optimal for tracking the deformation velocity peak position. The results depicted in Figure 10 correspond to the high-speed recording presented in Figure 9.

The particle velocity maps indicate that the shear wave propagation speed remains constant. This allows us to estimate $V_s$ by fitting a line to the maximum amplitude points in the map for each of the two "branches"—one for the negative positions and one for the positive positions. In the case of the bullet jet shown in Figure 10, $V_s$ is determined as $0.91 \pm 0.01$ m/s from the PIV data and $0.89 \pm 0.03$ m/s from the USI. Furthermore, the decay rate of the particle velocity could be also extracted from these kind of maps. Figure 10(c) depicts an example of the displacement caused by the passage of the shear wave through the outlined section of the gelatin block (as shown in the inset). The plot also displays the temporal evolution of particle velocity extracted from the ROI framed in red in both the optical (a) and the US displacement (b) maps of Figure 10. The difference in the width of the particle velocity peak obtained from the optical images and the USI indicates that the PIV method has a higher spatial resolution than the acoustic analogue. Here, we reduced the PIV interrogation window size until the optical results converged.

These measurements demonstrate that even a relatively small number of bubbles in the gelatin suffice to characterize shear wave dynamics and track its speed using ultrasound. Here, we had around 5 bubbles/mm$^2$ with a average cavity radius of 76.3 µm. The same processing method, as implemented in Figure 10, was applied to all available measurements. The mean propagation speeds for each jet type were as follows: $V_{sj} = 0.95 \pm 0.5$ m/s for the standard jet case, $V_{ij} = 0.92 \pm 0.5$ m/s for the intermediate jet case, and $V_{bj} = 0.96 \pm 0.5$ m/s for the bullet jet case. These values are consistent with those documented in other similar studies[46,47].

In our experiments, the elasticity modulus of the samples remained constant after an initial decay caused by the addition of water to the surface of the soft material block[33]. Measurements taken shortly after applying the water layer yielded an average speed of approximately 1.25 m/s, which using gelatin physical properties of $\rho = 1010$ kgm$^3$ and $\nu = 0.45$[48], results in $E = 4.5$ kPa. After approximately 40 minutes with water on top of the gelatin block, the absorption of liquid in the superficial layers of the samples slows considerably, and the measured shear wave speeds reach a plateau around 0.95 m/s. This translates to an elasticity modulus of 2.6 kPa. It is worth noting that the values of $V_s$ and consequently of the elasticity modulus can fluctuate significantly depending on the room temperature and gelatin curation method, but also with ratio between the shear wave amplitude and its distance to the rigid cuvette walls[46].

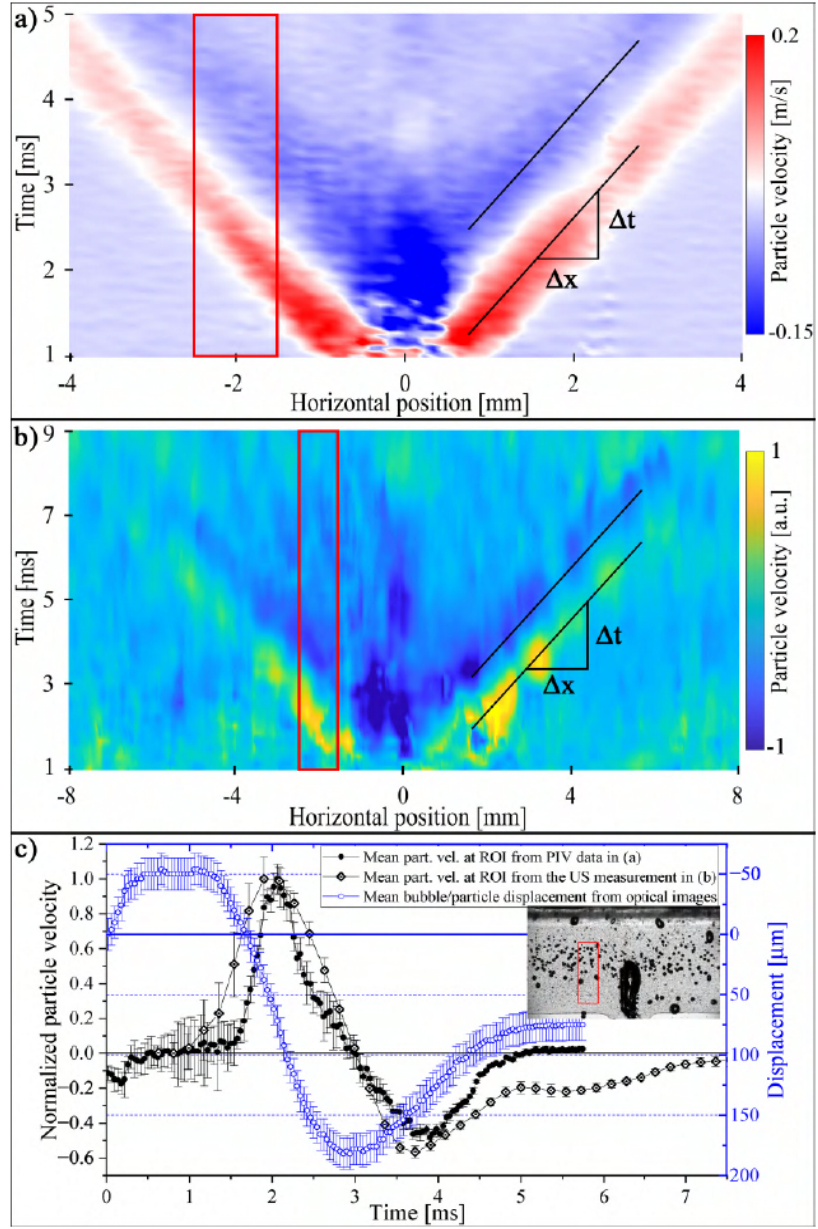

Figure 10. Particle velocity map (vertical component) as a function of horizontal position and time, averaged over a fixed window at the depth between 1.7 and 4.7 mm measured from the gelatine surface. A positive velocity indicates downward movement, $x = 0$ refers to bubble center in the horizontal direction and $t = 0$ is chosen as the bubble nucleation time. a) Map obtained from PIV applied to high-speed optical images with a spatial resolution of 25.3 µm pixel size and a temporal resolution of 40 µs. The high-velocity region shown in red illustrates the shear wavefront, followed by a negative velocity (shown in blue). b) Map obtained from the axial velocity map derived from the Loupas 2-D autocorrelator operated on the ultrasound data. Here we show normalized velocity field. The propagation speed of the shear wave was determined by fitting a line to the maximum displacement speed observed in the particle velocity maps, obtaining $V_s = 0.91 \pm 0.01$ m/s from the PIV method and $V_s = 0.89 \pm 0.03$ m/s from the USI. c) Velocity field, averaged over the $x$-coordinate withing a window defined between 1.5 and 2.5 mm shown in the red frames of a) and b). The evaluated region is also depicted in the inset of c).

## IV. CONCLUSIONS

We have investigated the potential of high-speed ultrasound imaging for studying fast events in multi-phase flows, gas diffusion through a gas/liquid interface and the propagation of shear waves in soft materials. This experimental method can operate in opaque materials. However, its effectiveness relies on the presence of acoustic scatterers within the material under examination. In this study, the scatterers were generated by laser-induced bubbles. Yet, these gas cavities can also be introduced directly into a liquid through alternative methods, such as a pressurized gas nozzle, hydrodynamic cavitation (e.g. liquid stirring) or by inducing boiling in the liquid.

In the initial experiments outlined in Section III A, we demonstrated the capability of compounding USI to characterize the flow induced by a jetting bubble penetrating into a pool of water. The frame rate of 5.5 kfps achieved by USI formed by three plane waves was enough to effectively characterized the complex resultant flow by tracking the surrounding gas injected through the free surface into the water.

Regarding the limitations of the method, we found that establishing a direct correlation between the intensity observed in the reconstructed USI and the actual size of the bubbles is not feasible, as determined from optical video measurements. The main source of inaccuracy in the ultrasound visualization technique is attributed to the "shielding" effect, particularly noticeable when bubbles group in small clusters. This effect causes them to appear larger than their actual size in the imaging. Furthermore, we observed that bubbles situated in close proximity to the free surface exhibit significantly higher relative intensity in the USI. This heightened intensity results from the robust reflection of ultrasound signals emitted by the probe on the liquid's free surface. On the contrary, bubbles at a close distance from the probe surface appear fainter in the USI. Consequently, in numerous instances, the system fails to detect these bubbles, leading to a dramatic reduction in the precision of the visualization method.

In spite of the limitations discussed above, the results presented in this study demonstrate the potential of ultrasound compounding images in investigating complex microfluidic flows beyond the usual medical applications, which are typically studied at a lower frame rate, e.g., below 2 kfps[20,22–24].

The detection of sub-micron bubbles, which were not observable optically, motivated the second experiment on the dissolution dynamics of laser-induced nanobubbles presented in Section III B. There, we established a strong correlation between the stability of bulk nanobubbles and the signals detected by the ultrasound transducer using a single plane wave mode. As these bubbles remained invisible to optical microscopy even for the smallest resolutions, their size distribution was taken from previous works employing the same generation technique[27,28], identifying bubbles with typical radius below 100 nm. These earlier measurements involved hundreds of repetitions of equivalent samples to estimate the characteristics of the bubble population. Other methods, such as dynamic light scattering, cannot differentiate bubbles from pollutants, such as solid particles or oil drops. Considering this, the current technique employing USI could represent a significant advancement in the field of nanobubbles[49,50]. High-speed USI enables direct detection of the void particles in a single-shot measurement, ensuring the gaseous nature of the detected objects. Nonetheless, the aforementioned limitations, particularly the inability to establish the minimum detectable bubble size, both with and without clustering, hinder our ability to perform an accurate analysis of the nanobubble sizes. Addressing these unknowns will demand further investigations.

In the third and last experiment, detailed in Section III C, we showcase the generation of shear waves within a soft material using a laser pulse, i.e., without mechanically accessing the samples. Subsequently, we successfully employ optical PIV and high-speed USI to studying the material's properties through shear wave elastography, finding an excellent agreement between both methods. From these measurements, it became clear that the "bullet" jets, characterized by a higher penetration, result in stronger wave fronts with a regular shape perpendicular to the jetting, differing from other jet types, where the propagation is predominantly spherical. Regarding the shear wave propagation speed ($V_s$), no significant variation was observed among the different jet types. This outcome was expected, given that the material's elasticity and density remained unaltered. Remarkably, this novel technique is capable of generating a stronger shear wave compared to previous methods that involved directly seeding the laser bubble inside the soft material[14,21,32]. The current approach also enables real-time evaluation of the effects of chemical reactions or biological processes on the samples, or changes due to humidity, temperature, and pressure in the sample.

An essential step in the preparation of the samples was adding homogeneously distributed bubbles to the bulk of the gelatin through laser energy deposition, serving as ultrasound contrast agents. This laser technique could also be used to seed an arbitrary localized distribution of cavities, for example, to study the propagation of shock and rarefaction waves, as performed by Ohl *et al.*[6], or to modify the mechanical properties of the soft material in a give region. Still, the bubble seeding process requires specific chemical and rheological conditions in the material to promote the growth and stabilization of the laser cavities[42].

## ACKNOWLEDGEMENTS

This project has received funding from the European Union's Horizon research and innovation programme under the Marie Skłodowska-Curie grant agreements No. 101064097 and No. 813766. It was also partly funded by the Federal Ministry of Education and the DFG (German Research Association) under Grant No. ME 1645/12-1.

## AUTHOR DECLARATIONS

### CONFLICT OF INTEREST

The authors have no conflicts to disclose.

### AUTHOR CONTRIBUTIONS

**Juan Manuel Rosselló**: Conceptualisation, Data curation, Formal Analysis, Funding acquisition, Investigation, Methodology, Visualisation and Writing – original draft, Writing – review & editing. **Saber Izak Ghasemian**: Conceptualisation, Data curation, Formal Analysis, Investigation, Software, Visualisation and Writing – original draft. **Claus-Dieter Ohl**: Conceptualisation, Funding acquisition, Resources and Writing – review & editing.